\documentclass[showpacs,aps,prb,preprint,floatfix,superscriptaddress]{revtex4}
\begin{document}
\title{Exchange-correlation vector potentials and
  vorticity-dependent exchange-correlation energy densities in
  two-dimensional systems} 
\author{Andreas Wensauer}
\author{Ulrich R\"ossler}
\affiliation{Institut f\"ur Theoretische Physik, 
         Universit\"at Regensburg, D-93040 Regensburg, Germany}
\date{\today}

\begin{abstract}

We present a new approach how to calculate the scalar
exchange-correlation potentials and the vector exchange-correlation
potentials from current-carrying 
ground states of two-dimensional quantum dots. From these
exchange-correlation potentials we derive exchange-correlation
energy densities and examine their
vorticity (or current) dependence. Compared with
parameterizations of current-induced effects 
in literature we find an increased significance
of corrections due to paramagnetic current densities. 

\end{abstract}

 
\maketitle

\section{Introduction}

Current-spin density functional theory (CSDFT)\cite{vignale87,
  vignale88} is a powerful tool to 
calculate ground state (GS) properties of interacting electron
systems. In contrast to spin-density functional theory
  (SDFT)\cite{barth72, rajagopal73} and
density-functional theory (DFT)\cite{hohenberg64, kohn65}, CSDFT
  takes into account the coupling of the magnetic 
field to both, the spin density and the magnetic current
  density. However, CSDFT calculations require the exchange 
  correlation (XC) energy as an input which is a functional of the
  spin densities $n_\sigma({\bf r})$ and the paramagnetic current
  density ${\bf j}_p({\bf r})$. In the local vorticity spin-density 
  approximation (LVSDA) the XC functional can be approximately
  calculated using XC energy densities from homogeneous systems
  which depend on the density parameter $r_s$, the polarization
  $\xi$ and the vorticity 
  ${\bf v}$ (or alternatively the local filling factor $f$). Having 
  investigated the XC energy density for two-dimensional (2D)
  systems with vanishing 
  current density in a previous work\cite{wensauer03d} the 
  focus of this paper is on current-induced effects on the XC energy
  density. The relation\cite{vignale87, vignale88} ${\bf v}=-e{\bf B}_{\rm
  hom}/m^\ast$ allows to link the vorticity ${\bf 
  v}=\nabla\times\frac{{\bf j}_p({\bf r})}{n({\bf r})}$ to
  homogeneous systems with a 
  fictitious magnetic field ${\bf B}_{\rm hom}$. This property is
  used by the current- or vorticity-induced corrections of XC energy
  densities in literature. They are parameterized as function of
  ${\bf B}_{\rm hom}$ or the local filling factor $f=2\pi n l_{\rm
  B}^2$. The local filling factor defines the number of occupied
  Landau levels of a 2D system in a magnetic field $B$. $l_{\rm
  B}=\sqrt{1/(e|{\bf B}_{\rm hom}|)}$ is the magnetic length
  corresponding to  ${\bf B}_{\rm hom}$ ($\hbar=1$ throughout this
  paper). 

CSDFT calculations for 2D systems (for quantum dots (QDs) see
e.g.\ Refs.\ [\onlinecite{ferconi94, steffens98a, reimann99}])
have made use of current-density dependent XC energy densities
based on interpolations provided by Levesque, Weis, and
MacDonald (LWM)\cite{levesque84} or Fano and Ortolani
(FO)\cite{fano88}. Both parameterizations are
based on the assumption that 
the coupling of the magnetic field to the currents is only relevant
for high $B$. For these strong magnetic fields the considered
system is in the lowest Landau level as the inter-Landau level
spacings increases with growing $B$ and the Zeeman energy and exchange
effects favor spin polarization.

For this regime Laughlin\cite{laughlin83} postulated the relation 
\begin{equation}
\varepsilon_{\rm{XC}}(f=m^{-1})={e^2\over l_{\rm B}\sqrt{2m}}\int_0^\infty
\!{\rm d} x\, [g_f(x l_{\rm B})-1]
\end{equation}
for filling factors $f=m^{-1}<1$ ($m$ is an odd number). 
For discrete filling factors $f={1\over 3},\, {1\over 5}$ Levesque,
Weis, and MacDonald (LWM)\cite{levesque84} calculate the XC
energy density by using the pair correlation function $g_f(r)$ of a
classical plasma with a density $n=f/(2\pi l_{\rm B}^2)$ and
statistical methods.\cite{caillol82} Interpolation delivers a
continuous expression for the dependence of the XC energy density on
the filling factor $f$ which is valid for $f<1$
\begin{equation}
\varepsilon_{\rm{XC}}^{\rm{LWM}}(r_s,f)/\mbox{Ry}= -0.782\, 133\,
{2\sqrt{2}\over r_s} \left( 1-0.211\,f^{0.74} + 0.012\,f^{1.7}\right).
\label{excLWM}
\end{equation}

Fano and Ortolani (FO)\cite{fano88} pursue an alternative concept to
parameterize current corrections for $\varepsilon_{\rm{XC}}$: They
follow the particle-hole symmetry for electrons in the lowest Landau
level at large magnetic fields and establish an interpolation which
exactly reproduces the well-known analytical result for filling
factor $f=1$ (see Ref.\ [\onlinecite{yoshioka83}]) 
\begin{eqnarray}
\varepsilon_{\rm{XC}}^{\rm{FO}}(r_s,f)/\mbox{Ry}&=& {2\sqrt{2}\over
f^{3/2} r_s}  \left[ -\sqrt{\pi\over 8}f^2-0.782\, 133\,f^{3/2}
\left(1-f\right)^{3/2}\right. \nonumber\\
&& + \left.
0.683\,f^2\left(1-f\right)^2-0.806\,f^{5/2}\left(1-f\right)^{5/2}
\rule{0mm}{6mm}
\right].
\label{FO}
\end{eqnarray}

In spite of concerns that the exact XC energy density might be
irregular and not differentiable,\cite{vignale95b}
the parameterizations (\ref{excLWM}) or (\ref{FO}) are used in a number of
papers\cite{ferconi94,reimann99,steffens98a,steffens_phd99}
assuming that they reproduce an important part of
current-induced XC effects. 
However, the LWM and the FO parameterization are only
correct in the high density limit, i.e.\ for  $r_s\rightarrow
0$.\cite{steffens_phd99} 

To obtain XC energy densities $\varepsilon_{\rm XC}(r_s,\xi,f)$
these current-dependent corrections are combined with XC
energy densities $\varepsilon_{\rm XC}^{\rm TC/LWM/FO}(r_s,\xi)$ from systems
without current-density by an interpolation established by Rasolt
and Perrot\cite{rasolt92}  
\begin{equation}
\varepsilon_{\rm{XC}}(r_s,\xi,f)=
{\varepsilon_{\rm{XC}}^{\rm{LWM/FO}}(r_s,f) + f^p 
\varepsilon_{\rm{XC}}^{\rm{TC/AMBG/ISI}}(r_s,\xi)\over 1+f^p}.
\label{interp}
\end{equation}
For large filling factors $f$ (i.e.\ for small magnetic
fields ${\bf B}_{\rm hom}$) the Pad\'e approximation yields the
well-known result of 
local spin density approximation (LSDA),
$\varepsilon_{\rm{XC}}^{\rm{TC/AMGB/ISI}}(r_s,\xi)$. 
For this quantity the interpolation by
Tanatar and Ceperley (TC)\cite{tanatar89} is the most frequently used
approach. However, recently new parameterizations were
proposed by Attaccalite, Moroni, Gori-Giorgi, and
Bachelet (AMGB)\cite{attaccalite02} and Seidl
(ISI=interaction strength interpolation)\cite{seidl01}. 
In the limit of
high effective magnetic fields or $f<1$ Eq.\ (\ref{interp}) takes
into account corrections due to paramagnetic currents. Usually the
exponent is chosen to be $p=4$ (see Ref.\
[\onlinecite{rasolt92}]). However, results seem to be 
almost insensitive with respect a variation of $p$.\cite{rasolt92,
  steffens_phd99} 
The usage of the interpolation (\ref{interp}) is justified by
an improvement of CSDFT results compared with benchmark calculations
from ED and Quantum-Monte-Carlo.\cite{ferconi94, steffens_phd99} 

It is also reported that results do not significantly depend on
whether the LWM or the FO parameterization is
applied.\cite{steffens_phd99} Even working with corrections which take
into account discontinuities in the derivative of the XC energy density with
respect to $f$ in the fractional quantum Hall regime (see
Ref.\ [\onlinecite{price96}]) or an additional 
dependence on the polarization (see Refs.\ [\onlinecite{oliver79,
  lubin97}]) do not further improve the CSDFT results.\cite{steffens_phd99} 

The present paper is a natural extension of our previous
publication\cite{wensauer03d} where we presented 2D XC energy densities
for systems without paramagnetic current density which were
extracted from exact GS densities 
and energies provided by exact diagonalization (ED)\cite{wensauer03b}. 
Here we want to focus on the current-induced effects on XC energy
densities. Analogous to Ref.\ [\onlinecite{wensauer03d}] the
extraction of XC energy densities is performed in two steps:
After briefly introducing CSDFT in Sec.\ II and the
properties of the QD Hamiltonian in Sec.\ III we present the first
step of the method, i.e.\ the calculation of scalar and vector XC
potentials from exact QD GSs together with results (Sec.\ IV). In
the following Section we introduce the second step which comprises the
concept how to obtain XC energy densities from scalar and vector XC
potentials. The results are compared with the current-dependent XC
energy densities discussed above. Sec.\ VI gives a summary of the
paper.   

\section{Current-spin-density functional theory and local vorticity
  (spin-)density approximation} 

In this Section we sketch the basics of CSDFT and LVSDA. 
The DFT formalism was originally established by
Hohenberg, Kohn, and Sham\cite{hohenberg64, kohn65} and generalized to 
spin-polarized systems\cite{barth72} and current-carrying
systems\cite{vignale87, vignale88} by including the
coupling of the polarization to an applied magnetic field and to the
paramagnetic current density. Accordingly,
the Hohenberg-Kohn (HK) theorem has to be modified with respect
to the spin degrees of freedom\cite{barth72} and to the paramagnetic
current density\cite{vignale87, vignale88}. For the most general
case, it states that two 
different non-degenerate GS wavefunctions $|\Psi\rangle$ and
$|\Psi^\prime\rangle$ always yield different combinations $(n_\sigma({\bf
  r}), {\bf j}_p({\bf r}))\neq (n^\prime_\sigma({\bf r}), {\bf
  j}^\prime_p({\bf r}))$ of spin densities and paramagnetic current
densities. This is sufficient to establish a functional of the total  
energy with the usual functional properties
\begin{equation}
E_{V_{0,\sigma}}[n_\sigma,{\bf j}_p]=F_{\rm HK}[n_\sigma,{\bf j}_p]+\sum_\sigma\int{\rm d}{\bf
  r}V_\sigma({\bf r})n_\sigma({\bf r})+e\int {\rm d}{\bf r}\, {\bf
  j}_p({\bf r})\cdot {\bf A}({\bf r})+\frac{e^2}{2m^\ast}\sum_\sigma
  \int {\rm d}{\bf r}\, n_\sigma({\bf r}){\bf A}^2({\bf r})
\end{equation}
and the universal HK functional 
\begin{equation}
F_{\rm HK}[n_\sigma,{\bf j}_p]=\langle\Psi[n_\sigma,{\bf j}_p]|T+W|\Psi[n_\sigma,{\bf j}_p]\rangle .
\end{equation}

However, in contrast to the original HK theorem\cite{hohenberg64} the
uniqueness of the potentials is lost in CSDFT, because it is
possible that two different combinations of spin-dependent
potentials and vector potential yield the
same GS wavefunction. Thus, the one-to-one map between GS densities
and XC potentials is destroyed. The consequence is that there
might be discontinuities in the functional derivatives of the HK
functional or the XC energy functional with respect to the
densities.\cite{capelle02} Capelle and
Vignale\cite{capelle02} conclude that this problem also affects the
extraction of exact XC potentials from given GS densities. However,
in the case of axially symmetric, parabolic quantum dots it is
possible to restore the uniqueness as we will see in Section \ref{n2v}.

For practical purposes the variational scheme has to be mapped to
the Kohn-Sham (KS) system, i.e.\ an effective single-particle system with
the same GS densities as the interacting system.
The spin-degree of freedom is considered in the KS equations\cite{barth72} by
assuming the total spin $S_z$ in $z$-direction to be a good quantum number
\begin{eqnarray}
\left[{1\over 2m^\ast}\left\{{\nabla\over {\rm i}}
+e\left[{\bf A}({\bf r})+{\bf A}_{\rm{XC}}({\bf r})\right]
\right\}^2 +{e^2\over
2m^\ast}\right. \left.\left\{A^2(\bf{r})-\left[\bf{A}(\bf{r})+
{\bf A}_{\rm{XC}}({\bf r})\right]^2\right\}\right.&&\nonumber\\
 \left.\phantom{\left\{{1\over 2m^\ast}\right\}^2}+
V_\sigma({\bf r})+V_{\rm{H}}({\bf r})
+V_{\rm{XC},\sigma}({\bf r})
\right]\varphi_{j,\sigma}({\bf r})
&=& \varepsilon_{j,\sigma}\varphi_{j,\sigma}({\bf r})
\label{kseq}
\end{eqnarray}
with the spin $\sigma=\pm$ in $z$-direction and the KS energies
$\varepsilon_{j,\sigma}$ in increasing order. For a
system containing $N$ particles 
the spin densities and the paramagnetic current density are given by
\begin{eqnarray}
n_\sigma({\bf r})&=&
\sum_j\gamma_{j,\sigma}\left|\varphi_{j,\sigma}({\bf r})\right|^2\\
j_p({\bf r})&=&\sum_{j,\sigma}\gamma_{j,\sigma}\left[
\varphi_{j,\sigma}^\ast({\bf r})\nabla\varphi_{j,\sigma}({\bf r})-
\left(\nabla\varphi_{j,\sigma}^\ast({\bf r})\right)\varphi_{j,\sigma}({\bf r})
\right] 
\end{eqnarray}
with $\gamma_{j,\sigma}$ being occupation numbers
of the KS levels in the GS 
($\sum_j \gamma_{j,\sigma}=N_\sigma$ and $N_++N_-=N$).
Then the GS density and polarization are
\begin{eqnarray}
n({\bf r})&=&n_+({\bf r})+n_-({\bf r})\\
\xi({\bf r})&=&\frac{n_+({\bf r})-n_-({\bf r})}{n({\bf r})}
\end{eqnarray}
The XC potentials 
\begin{eqnarray}
V_{\rm XC,\sigma}([n_\sigma,{\bf j}_p],{\bf r})&=&
\frac{\delta E_{\rm XC}[n_\sigma,{\bf j}_p]}{\delta n_\sigma({\bf
    r})}\label{vxcdef}\\
eA_{\rm XC,\sigma}([n_\sigma,{\bf j}_p],{\bf r})&=&
\frac{\delta E_{\rm XC}[n_\sigma,{\bf j}_p]}{\delta {\bf j}_p({\bf
    r})}\label{axcdef}
\end{eqnarray}
are defined as functional
derivatives of the XC energy functional
\begin{equation}
E_{\rm XC}[n_\sigma,{\bf j}_p]=F_{\rm HK}[n_\sigma,{\bf j}_p]-
\frac{1}{2}\frac{e^2}{4\pi\varepsilon\varepsilon_0}
\int{\rm d}{\bf r}\int{\rm d}{\bf r}^\prime
\frac{n({\bf r})n({\bf r}^\prime)}
{\left|{\bf r}-{\bf r}^\prime\right|}-T_{\rm S}[n_\sigma,{\bf j}_p].
\end{equation}
($T_{\rm S}[n_\sigma,{\bf j}_p]$ denotes the kinetic energy functional of the KS
system.)
The total GS energy $E_0$ of the interacting system can be calculated
from 
\begin{eqnarray}
E_0&=&\sum_{j,\sigma}\gamma_{j,\sigma}\varepsilon_{j,\sigma}-
\frac{1}{2}\frac{e^2}{4\pi\varepsilon\varepsilon_0}
\int{\rm d}{\bf r}\int{\rm d}{\bf r}^\prime
\frac{n({\bf r})n({\bf r}^\prime)}
{\left|{\bf r}-{\bf r}^\prime\right|}
-\sum_\sigma\int{\rm d}{\bf r}\,V_{\rm XC,\sigma}
([n_\sigma,{\bf j}_p],{\bf r})n_\sigma({\bf r})\nonumber\\
&&-e\int{\rm d}{\bf r}\,{\bf j}_p({\bf r})\cdot{\bf A}_{\rm XC}({\bf r})
+E_{\rm XC}[n_\sigma,{\bf j}_p]. 
\label{Eformel}
\end{eqnarray}
Concerning the XC potentials we apply the LVSDA 
\begin{equation}
E_{\rm XC}[n_\sigma,{\bf v}]\approx
\int{\rm d}{\bf r}\,n({\bf r})
\varepsilon_{\rm XC}(n_\sigma({\bf r}),{\bf v}({\bf r})).
\label{lvsdaeq}
\end{equation}
Using the vorticity ${\bf v}({\bf r})=\nabla\times\frac{{\bf
    j}_p({\bf r})}{n({\bf r})}$ 
instead of the paramagnetic current density is a consequence of the
gauge invariance of CSDFT and appropriate for local density
approximations.\cite{vignale87,vignale88} 
The most important parameterizations of the XC energy density
used in 2D calculations were introduced in Sec.\ I. 

\section{Quantum dot Hamiltonian and ground state densities}

We consider a two-dimensional QD with an axially symmetric
parabolic confinement potential of strength $\omega_0$
in a magnetic field ${\bf B}=(0,0,B)$. 
The Hamiltonian for $N$ particles in real-space representation (with
${\bf r}=(x,y)$, ${\bf p}= (p_x,p_y)$, angular momentum in
$z$-direction $l_z=xp_y-yp_x=\frac{1}{i}\frac{\partial}{\partial
\varphi}$, and vector potential ${\bf
  A}({\bf r})=\frac{B}{2}(-y,x)$) reads  
\begin{eqnarray}
 H=\sum_{j=1}^N \left(\frac{1}{2m^\ast}\left({\bf p}_j+e{\bf
      A}({\bf r}_j) \right)^2 +
\frac{1}{2}m^\ast\omega_0^2{\bf r}_j^2 \right)
+\frac{1}{2} \sum_{j,k=1}^{N}{\!\! ^\prime}\,\,
\frac{e^2}{4\pi\varepsilon\varepsilon_0|{\bf r}_j-{\bf r}_k|}
\label{ham_rss}
\end{eqnarray}
or 
\begin{eqnarray}
 H=\sum_{j=1}^N \left(\frac{{\bf p}_j^2}{2m^\ast} +
\frac{1}{2}m^\ast\omega_h^2{\bf r}_j^2 +\frac{\omega_c}{2}l_z\right)
+\frac{1}{2} \sum_{j,k=1}^{N}{\!\! ^\prime}\,\,
\frac{e^2}{4\pi\varepsilon\varepsilon_0|{\bf r}_j-{\bf r}_k|}.
\label{ham_rss2}
\end{eqnarray}
Here $e$ is the electron charge, $m^\ast=0.067\, m_e$ and 
$\varepsilon=12.4$ are the effective mass and the screening constant
of the host semiconductor (assumed to be GaAs),
$\omega_c=\frac{eB}{m^\ast}$ is the cyclotron frequency, and
$\omega_h=\sqrt{\omega_0^2+\omega_c^2/4}$ is
the hybrid frequency.

The reference densities and energies for the
investigation of XC energy densities are calculated by ED techniques which
provide results of high accuracy.\cite{wensauer03b} 
Note that the (spin-)density and the current density of all
eigenstates of the angular 
momentum operator are functions of radius $r$ but not of the angle
$\varphi$. Moreover, the current density is always of the form ${\bf
  j}_p({\bf r})=j_{p,\varphi}(r)\,{\bf e}_\varphi$ whereas the
radial component vanishes. 
Therefore, the relevant quantities provided by ED are the
spin-densities $n_\sigma(r)$ and the azimuthal component
$j_{p,\varphi}(r)$ of the
paramagnetic current density (or the $z$-component $v$ of
the vorticity) of the GS and its energy $E_0$.   

The many-particle Hamiltonian also shows an interesting scaling
property\cite{wensauer03b} which 
plays an important role in the context of the HK theorem.
It is revealed if we rewrite (\ref{ham_rss2}) as a function of $\omega_0$,
$\omega_c$, and $N$ and separate the angular momentum contribution. 
This concept is similar to the modification of the Hamiltonian of a
(natural) atom in Ref.\ [\onlinecite{capelle02}]:
\begin{eqnarray}
H(\omega_0,\omega_{\rm c},N)&=&\sum_{j=1}^N \left(\frac{1}{2m^\ast}{\bf p}_j^2+
\frac{1}{2}m^\ast\left(\omega_0^2+\frac{\omega_c^2}{4}\right){\bf r}_j^2\right)
+\frac{1}{2} \sum_{j,k=1}^{N} {\!\! ^\prime}\,\,
\frac{e^2}{4\pi\varepsilon\varepsilon_0|{\bf r}_j-{\bf r}_k|}
+\sum_{j,k=1}^{N}\frac{\omega_c}{2} l_j
\nonumber\\
&=& H(\sqrt{\omega_0^2+\omega_c^2/4},0,N)+\frac{\omega_c}{2} L.
\label{scaling2}
\end{eqnarray}
We identify the first part of (\ref{scaling2}) as a
Hamiltonian of a QD in zero magnetic field with the confinement frequency
$\sqrt{\omega_0^2+\omega_c^2/4}$. The sum over all single-particle angular
momenta in the last term yields the total angular momentum. As a result
we can map the spectra and wave functions of QDs in the magnetic field
to QDs in zero magnetic field.

The KS Hamiltonian of axially symmetric 2D QDs reads
\begin{equation}
H_{{\rm S},\sigma}=-\frac{1}{2m^\ast}\left\lbrack
  \frac{1}{r}\frac{\partial}{\partial
  r}\left(r\frac{\partial}{\partial r} \right)+\frac{1}{r^2}\frac{\partial^2}{\partial\varphi^2} \right\rbrack
  +\frac{1}{2}m^\ast\omega_0^2r^2+V_{\rm H}(r)+V_{\rm
  XC,\sigma}([n_\sigma,v],r) 
\label{effksham}
\end{equation}
Obviously, the KS Hamiltonian shows the same scaling properties as
the original Hamiltonian
\begin{equation}
H_{{\rm S},\sigma}(\omega_0,\omega_{\rm c},N)=
H_{{\rm S},\sigma}(\sqrt{\omega_0^2+\omega_c^2/4},0,N)+\frac{\omega_c}{2} L. 
\label{scalingks2}
\end{equation}
Therefore, the XC effects do not depend on the strength of the
external magnetic field. We can restrict our studies to 
systems without magnetic field. However, without magnetic field we
can apply the original HK theorem to the Hamiltonian (\ref{ham_rss})
what guarantees the uniqueness of the wavefunction and of the external
potential. For the KS system the uniqueness of the (XC) potentials
in quantum dots is explicitly investigated in the following Section.  

\section{Scalar and vector XC potentials}
\label{n2v}

A review of the literature on the numerical inversion of KS
equations, i.e.\ the 
calculation of XC potentials from GS densities can be found in
Ref.~[\onlinecite{wensauer03d}]. However, in this paper we will focus on
an unpolarized and on a fully polarized system  for which we can
analytically calculate the scalar XC 
potential and the XC vector potential. Before we present our results
for $V_{\rm XC}$ and $A_{\rm XC}$ we will discuss the problem of
uniqueness of XC potentials in CSDFT in the next paragraph.

\subsection{Uniqueness of XC (vector) potentials}

Because of Eqs.\ (\ref{scaling2}) and (\ref{scalingks2}) we can
assume the single-particle states  
$|\varphi_{j,\sigma}\rangle$ ($j$ is a combination of radial and
angular momentum quantum number) of the KS Slater determinant 
\begin{equation}
|\Psi\rangle =|\varphi_{1,\downarrow}\rangle
...|\varphi_{N_\downarrow,\downarrow}\rangle
|\varphi_{1,\uparrow}\rangle ...|\varphi_{N_\uparrow,\uparrow}\rangle
\end{equation}
to be eigenfunctions of the KS Hamiltonian (\ref{effksham}) at $B=0$
\begin{eqnarray} 
 H_{\rm{KS}} &=& -{1\over 2m^\ast}\left[ {1\over r}{\partial\over\partial
r}\left( r{\partial\over\partial r}\right)+{1\over r^2}
{\partial^2\over \partial\varphi^2}\right] +{1\over 2}m^\ast
\omega_0^2r^2 +{\partial \over \partial\varphi}{e\over {\rm i}
  m^\ast} {A_{\rm XC,\varphi}(r)\over r} \nonumber \\ 
&&+  V_{\rm{H}}(r) + \sum_\sigma V_{\rm{XC}\,\sigma}(r)
|\sigma\rangle\langle\sigma|.
\label{ks}
\end{eqnarray}
with the potentials $( V_{\rm{XC}\,\sigma}, A_{\rm XC,\varphi})$ 
\begin{equation} 
 H_{\rm KS}|\varphi_{j,\sigma}\rangle=\varepsilon_{j,\sigma}|\varphi_{j,\sigma}\rangle.
\label{diff1}
\end{equation}
Due to the axial symmetry of the system the wavefunctions
$|\varphi_{j,\sigma}\rangle$ are also
eigenfunctions of the angular momentum operator
$ l|\varphi_{j,\sigma}\rangle=l_{j,\sigma}|\varphi_{j,\sigma}\rangle$.

Now let us assume that there is a KS Hamiltonian
$H^\prime_{\rm{KS}}$ with the potentials
$( V^\prime_{\rm{XC}\,\sigma}, A^\prime_{\rm XC,\varphi})\ne
( V_{\rm{XC}\,\sigma}+c_\sigma, A_{\rm XC,\varphi})$ so that
the KS single-particle wavefunctions are also eigenfunctions of
$H^\prime_{\rm{KS}}$ 
\begin{equation} 
 H^\prime_{\rm KS}|\varphi_{j,\sigma}\rangle=\varepsilon^\prime_{j,\sigma}|\varphi_{j,\sigma}\rangle.
\label{diff2}
\end{equation}
Subtracting Eq.\ (\ref{diff1}) from Eq.\ (\ref{diff2}) delivers the constraint
\begin{equation} 
\Delta V_{{\rm XC},\sigma}(r) +l_{j,\sigma} {e\over
m^\ast} {{\Delta A_{\rm XC,\varphi}(r)}\over
r}-\varepsilon^\prime_{j,\sigma}+\varepsilon_{j,\sigma}=0
\label{bed}
\end{equation}
for the potential differences $\Delta V_{{\rm
XC},\sigma}(r)=V^\prime_{{\rm XC},\sigma}(r)-V_{{\rm XC},\sigma}(r)$
and $\Delta 
A_{\rm XC,\varphi}(r)=A^\prime_{\rm XC,\varphi}(r)-A_{\rm XC,\varphi}(r)$.
Eq.\ (\ref{bed}) has to be fulfilled for all wavefunctions
$|\varphi_{j,\sigma}\rangle$.  

If at least two states with different angular momenta
$l_{j,\sigma}\ne l_{j^\prime,\sigma}$ are occupied we can conclude
from 
\begin{eqnarray} 
&&l_{j^\prime,\sigma}\left(\Delta V_{{\rm XC},\sigma}(r) +l_{j,\sigma} {e\over
m^\ast} {{\Delta A_{\rm XC,\varphi}(r)}\over
r}-\varepsilon^\prime_{j,\sigma}+\varepsilon_{j,\sigma}\right)\nonumber\\
&&-l_{j,\sigma}\left(\Delta V_{{\rm XC},\sigma}(r) +l_{j^\prime,\sigma} {e\over
m^\ast} {{\Delta A_{\rm XC,\varphi}(r)}\over
r}-\varepsilon^\prime_{j^\prime,\sigma}+\varepsilon_{j^\prime,\sigma}\right)=0,
\end{eqnarray}
that the scalar potentials agree up to a constant 
$\Delta V_{{\rm XC},\sigma}(r)={\rm const}$.
Thus, $\Delta A_{{\rm XC},\varphi}(r)={\rm const^\prime}\cdot r$
immediately follows. 
As a consequence of finite XC vector potentials at $B=0$ 
we derive $\Delta A_{{\rm XC},\varphi}(r)=0$. If there are also
occupied states in the other spin direction we obtain 
$\Delta V_{{\rm XC},-\sigma}(r)={\rm const}$.
This result is in contradiction to
$( V^\prime_{\rm{XC},\sigma}, A^\prime_{\rm XC,\varphi})\ne
( V_{\rm{XC}\,\sigma}+c_\sigma, A_{\rm XC,\varphi})$ and the
uniqueness of XC potentials is proven.

The assumption of at least two occupied states with different
angular momenta $l_{j,\sigma}\ne l_{j^\prime,\sigma}$ in one spin
direction is correct for the configurations considered in this
paper. All other cases are more or less pathological:
\begin{itemize}
\item
For $N_\uparrow=N_{\downarrow}=1$ uniqueness cannot be proven
except for the the case $l_{1\uparrow}=l_{1\downarrow}=0$ (i.e.\
conventional DFT without paramagnetic current densities and vanishing
$A_{\rm XC}$).

\item
In the case of $N \ge 2$,
$l_{1\downarrow}=...=l_{N_\downarrow\downarrow}$ and 
$l_{1\uparrow}=...=l_{N_\uparrow\uparrow}$, uniqueness cannot be
guaranteed either. However, such a configuration should not be a GS
in particular for larger particle numbers.
\end{itemize}

In general the XC potentials are overdetermined by
(\ref{bed}). Therefore, the crucial aspect is the existence of 
appropriate XC potentials but not their uniqueness.

\subsection{Results for XC (vector) potentials}

From the discussion of the uniqueness of XC potentials presented
above immediately follows that we can
analytically calculate the scalar and the vector potentials if we
are able to derive the exact KS wavefunctions from $(n_\sigma({\bf
  r}),j_p({\bf r}))$. This can be accomplished for special systems.

The simplest conceivable configuration is the fully polarized
two-electron $\nu=1$-quantum-Hall-droplet (QHD) with
quantum numbers $L=-1$,
$S=1$, and $S_z=-1$. In particular the confinement energy
$\omega_{\rm h}=2/3{\rm Ry}$ allows for a completely analytical
solution.\cite{taut94} (For the rest of this subsection we denote energies
in ${\rm Ry}$ and lengths in effective Bohr radii ${\rm
  a}_0$.) According to
Taut\cite{taut94} the separation of center-of-mass and relative
motion delivers the GS wavefunction
\begin{eqnarray}
\Psi({\bf r}_1,{\bf r}_1)&=&\varphi_{\rm CM}(|{\bf r}_1+{\bf
r}_2|/2)\varphi_{\rm rel}(|{\bf r}_1-{\bf r}_2|)\nonumber\\
&=&C_{\rm CM} {\rm e}^{-R^2/3}\,C_{\rm rel}
r\left(1+r/3\right){\rm e}^{-{r^2}/12}
\end{eqnarray}
(${\bf R}=({\bf r}_1+{\bf r}_2)/2)$ and ${\bf r}={\bf r}_1-{\bf r}_2$)
and GS energy $8/3$. The constants are $C_{\rm
  CM}={\sqrt{\frac{2}{3\pi}}}$ and $C_{\rm
  rel}=1/{\sqrt{2\pi\left(9\sqrt{6\pi}+42 \right)}}$. 

Using these results we can analytically calculate the GS density
$n(r)$, the azimuthal component of the paramagnetic current
density $j_{{p},\varphi}(r)$, and the ($z$-component of the) vorticity
\begin{equation}
v({\bf r})=\frac{1}{r}\frac{\partial}{\partial
r}\left(r\,\frac{j_{{p},\varphi}(r)}{n(r)}\right).
\end{equation}

Now we are interested in the effective potentials which exactly
reproduce the GS density combination $(n(r),j_{{\rm
    p},\varphi}(r))$. In this context we make use of the scaling
relations (\ref{scaling2}) and (\ref{scalingks2}) and restrict
ourselves to the case $B=0$.  

Labeling states and occupation numbers by three indices which
characterize the radial quantum number, the angular momentum and the
spin state, the GS of the (spin-down)
$\nu=1$-QHD in the KS system is given by
$\gamma_{0,0,-1}=\gamma_{0,-1,-1}=1$. Then the relations between 
KS wavefunctions $\phi_{j,l,\sigma}({\bf
r})=\frac{1}{\sqrt{2\pi}}{\rm e}^{{\rm i}m\varphi}R_{j,l}(r)$ and
density (Fig.~\ref{2el_ana}(a))
\begin{equation}
n({\bf r})=\sum_{j,l,\sigma} \gamma_{j,l,\sigma} \phi_{j,l,\sigma}^\ast({\bf
r})\phi_{j,l,\sigma}({\bf r}) 
\label{inversion_anfang}
\end{equation}
and paramagnetic current density (Fig.~\ref{2el_ana}(b))
\begin{equation}
j_{{p},\varphi}(r)=\frac{2\,{\rm Ry}}{r}
\sum_{j,l,\sigma} \gamma_{j,l,\sigma}l \phi_{j,l,\sigma}^\ast({\bf
r})\phi_{j,l,\sigma}({\bf r}) 
\end{equation}
allow for the calculation of the vorticity (Fig.~\ref{2el_ana}(c))
and of the exact KS orbitals 
\begin{equation}
R_{0,1}(r)=\sqrt{(\pi/{\rm Ry}) r j_{{p},\varphi}(r)}
\end{equation} 
and
\begin{equation}
R_{0,0}(r)=\sqrt{2\pi n(r)-R_{0,1}^2(r)}.
\end{equation}
The radial KS wavefunctions $R_{0,0}(r)$ und $R_{0,1}(r)$ are
depicted in Fig.~\ref{2el_ana}(d).

Consequently, we can calculate the effective scalar single-particle
potential up 
to a constant $\varepsilon_{0,0,-1}$ (which is a KS energy) by
inverting the KS Schr\"odinger equation for $\phi_{0,0,-1}({\bf r})$
(see Fig.~\ref{2el_ana}(e))
\begin{equation}
V_{\rm eff}(r)=r^2/9+V_{\rm H}(r)+V_{\rm XC}(r)-\varepsilon_{0,0,-1}=
\frac{\left[\frac{1}{r}\frac{\partial}{\partial
r}\left(r\frac{\partial}{\partial r} \right) \right] R_{0,0}(r)}{R_{0,0}(r)}.
\end{equation}
In particular the KS equation for states with $l=0$ only involves
the scalar potential, but not the XC vector potential.
As for given density $n(r)$ the Hartree-potential $V_{\rm H}(r)$
(Fig.~\ref{2el_ana}(f)) is
also known we can derive the exact scalar XC
potential (Fig.~\ref{2el_ana}(g))
\begin{equation}
V_{\rm XC}(r)=\varepsilon_{0,0,-1}-r^2/9-V_{\rm H}(r)
+\frac{\left[\frac{1}{r}\frac{\partial}{\partial
r}\left(r\frac{\partial}{\partial r} \right) \right] R_{0,0}(r)}{R_{0,0}(r)}
\end{equation}
up to the gauge constant $\varepsilon_{0,0,-1}$.

Finally we can find an approach to the XC vector potential
(Fig.~\ref{2el_ana}(h)) by exploiting the KS equation for
$\phi_{0,-1,-1}({\bf r})$. After solving the KS equation for
$A_{{\rm XC},\varphi}(r)$ and substituting the scalar XC potential
we obtain 
\begin{equation}
eA_{{\rm XC},\varphi}(r)=\frac{1}{\rm Ry}\frac{r}{2}
\left(
\frac{1}{r^2}
-\frac{\left[\frac{1}{r}\frac{\partial}{\partial
r}\left(r\frac{\partial}{\partial r} \right) \right] R_{0,1}(r)}{R_{0,1}(r)}
+\frac{\left[\frac{1}{r}\frac{\partial}{\partial
r}\left(r\frac{\partial}{\partial r} \right) \right] R_{0,0}(r)}{R_{0,0}(r)}
+\varepsilon_{0,0,-1}-\varepsilon_{0,-1,-1}
\right).
\label{axcalsdiff}
\end{equation}
The singularity in the term $1/r^2$ is
eliminated by the second one as both terms together represent the
kinetic energy density of the state $\phi_{0,-1,-1}({\bf r})$ which
is finite at $r=0$.

Now let us focus on the asymptotic of all relevant quantities. In
the limit of large radii the asymptotic expansions of the densities
and the vorticity are 
\begin{eqnarray}
&&\lim_{r\rightarrow\infty}n(r)\rightarrow\frac{2\pi}{3} C_{\rm CM}^2 C_{\rm
rel}^2 r^4\, {\rm e}^{-r^2/3}\\
&&\lim_{r\rightarrow\infty}j_{{p},\varphi}(r)\rightarrow\frac{4\pi C_{\rm
CM}^2 C_{\rm rel}^2}{3(\,1/{\rm Ry})}r^3 \,{\rm e}^{-r^2/3}  \\
&&\lim_{r\rightarrow\infty}v(r)\rightarrow{\rm Ry}\left(-24/r^4+54/r^5\right).
\end{eqnarray}
Thus, the results for the radial wavefunctions $R_{0,0}(r)$ and
$R_{0,1}(r)$ are
\begin{eqnarray}
&&\lim_{r\rightarrow\infty}R_{0,0}(r)\rightarrow2\sqrt{2}\pi C_{\rm CM} C_{\rm
rel} r\,{\rm e}^{-r^2/6}\\ 
&&\lim_{r\rightarrow\infty}R_{0,1}(r)\rightarrow\frac{2\pi C_{\rm CM} C_{\rm
rel}}{\sqrt{3}} r^2\,{\rm e}^{-r^2/6}. 
\end{eqnarray}
The effective potential is parabolic in the limit of large $r$
\begin{equation}
\lim_{r\rightarrow\infty}V_{\rm eff}(r)\rightarrow\frac{1}{9}r^2-\frac{4}{3}+\frac{3}{2r}+\frac{3}{4r^2}.
\end{equation}
With the Hartree potential converging to zero
\begin{equation}
\lim_{r\rightarrow\infty}V_{\rm H}(r)\rightarrow\frac{4}{r}
\end{equation}  
we assume that the XC potential vanishes in the limit
$r\rightarrow\infty$.\cite{laufer86} This is achieved by an gauge constant
$\varepsilon_{0,0,-1}=4/3$ 
\begin{equation}
\lim_{r\rightarrow\infty}V_{\rm XC}(r)\rightarrow-\frac{5}{2r}+\frac{3}{4r^2}.
\end{equation}  
The asymptotic of the XC vector potential gives a constraint for 
$\varepsilon_{0,-1,-1}=2$ if we do not want $A_{{\rm
    XC},\varphi}(r)$ to diverge in the limit of large radii. (This is
a reasonable requirement as we work at $B=0$. A contribution $A_{{\rm
XC},\varphi}(r)=\frac{B^\prime}{2}r$ would correspond to an external
magnetic field ${\bf B}^\prime=(0,0,B^\prime)$.)
\begin{equation}
\lim_{r\rightarrow\infty}A_{{\rm
XC},\varphi}(r)\rightarrow\frac{1}{\rm Ry}\left(-\frac{1}{4}-\frac{7}{8r}\right). 
\end{equation}  
Thus, the XC potentials $(V_{\rm XC}(r),A_{{\rm XC},\varphi}(r))$
are uniquely determined.

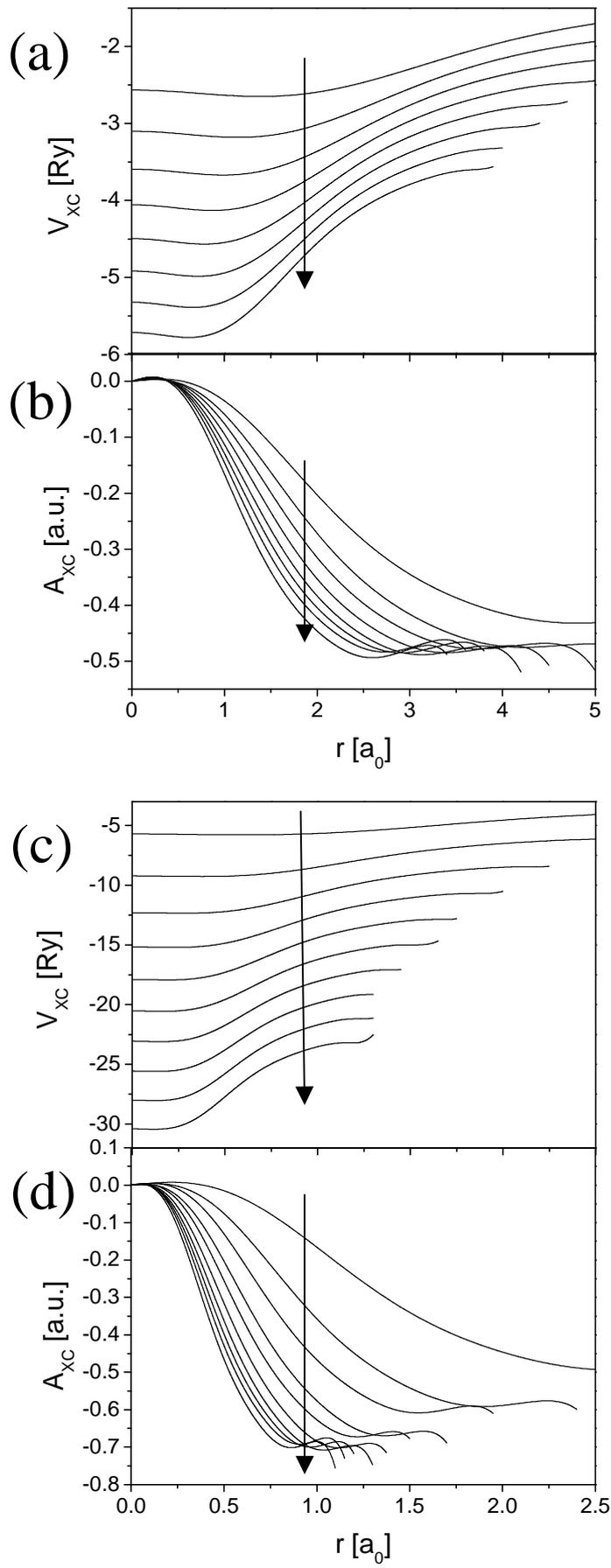
\begin{figure}
\caption{Scalar XC potentials (a) and (c) and XC vector potentials
  (b) and (d) of the two-electron $\nu=1$-QHD as a function
  of the strength of the confinement potential. In (a) and (b) the
  results for $\omega_0=3\,{\rm meV}$ - $10\,{\rm meV}$ (stepsize
  $1\,{\rm meV}$) are shown, in (c) and (d) results for
  $\omega_0=10\,{\rm meV}$ - $100\,{\rm meV}$ (stepsize 
  $10\,{\rm meV}$).
The arrows indicate curves corresponding to increasing confinement
potentials.  
\label{2el_varycon}
}
\end{figure}

Besides this fully analytic approach it is also possible to perform
the calculation of XC potentials starting with GS densities from
ED. Using numerical data for the densities allows us to study
the dependence of the XC vector potential on the strength
of the external confinement potential. In Fig.~\ref{2el_varycon} we show the
scalar XC potentials (up to a constant) and the XC vector potentials for
$\omega_0$ in the range between $3\,{\rm meV}$ and $100\,{\rm meV}$. 
In the context of the XC vector potentials we find a pronounced
sensitivity to the precision of the densities at the edge of the
dot. Because the $A_{{\rm XC},\varphi}(r))$ is calculated as a difference
of two (large) effective potentials, numerical errors are growing at
the edge of the dot although 16 Landau levels were used in
ED. However, the results from the bulk part of the QD are reliable
and show a continuous dependence on the strength of $\omega_0$. 

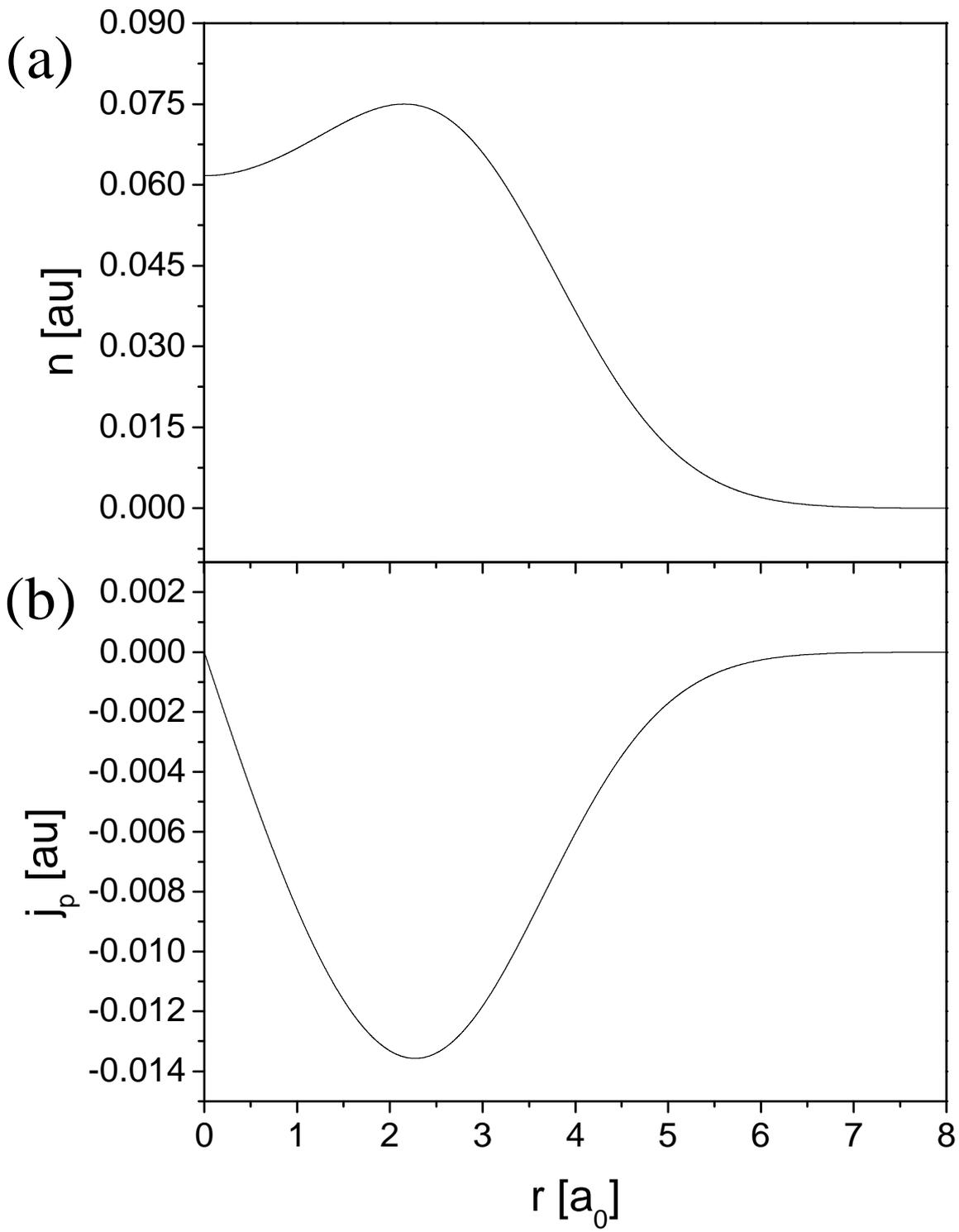
\begin{figure}
\caption{GS density (a) and GS paramagnetic current density (b) of a
  four-electron $\nu=2$-QHD with confinement energy $3\,{\rm meV}$.
\label{4el_example}
}
\end{figure}

The second system for which we study scalar and vector XC
potentials is the unpolarized four-electron system ($\nu=2$-QHD)
with $S=0$, $L=-2$. In this case all GS densities are provided
by ED using 335259 Slater determinants (see Fig.~\ref{4el_example} for GS
(paramagnetic current) densities at $\omega_0=3\,{\rm
  meV}$). Concerning the mathematical structure of the KS equations
it is similar to the $\nu=1$ two-electron droplet: the $\nu=2$
four-electron droplet can be regarded as a combination of two
$\nu=1$ two-electron droplets with different spin
polarizations. Therefore the inversion of KS equations is completely
analogous to Eqs.\ (\ref{inversion_anfang})-(\ref{axcalsdiff}) if we
use half of the GS 
density and paramagnetic density of the four-electron $\nu=2$-QHD.  

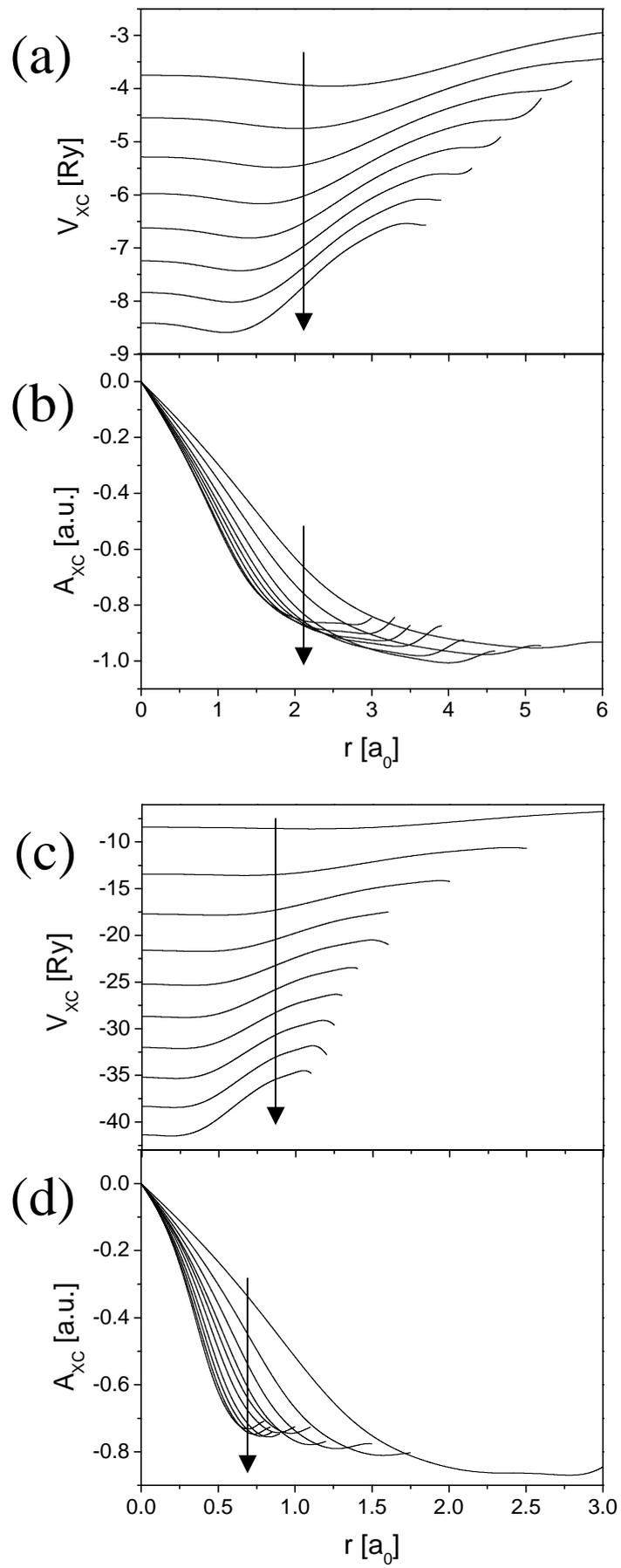
\begin{figure}
\caption{Scalar XC potentials (a) and (c) and XC vector potentials
  (b) and (d) of the four-electron $\nu=2$-QHD as a function
  of the strength of the confinement potential. In (a) and (b) the
  results for $\omega_0=3\,{\rm meV}$ - $10\,{\rm meV}$ (stepsize
  $1\,{\rm meV}$) are shown, in (c) and (d) results for
  $\omega_0=10\,{\rm meV}$ - $100\,{\rm meV}$ (stepsize 
  $10\,{\rm meV}$).
The arrows indicate curves corresponding to increasing confinement
potentials.
\label{4el_varycon}
}
\end{figure}

Fig.~\ref{4el_varycon} shows the results for scalar XC potentials
(up to a
constant) and vector XC potentials as a function of the strength of
the confinement potential. Concerning the precision of the vector
potentials we can repeat the same arguments as for the two-electron
system, i.e.\ only the values from the bulk part of the QD are
reliable. Similar to the $\nu=1$-droplet we find again a continuous
dependence on $\omega_0$. Another aspect is the size of the amplitude
of the XC vector potential which is even larger than for the two
electron system (see Fig.~\ref{2el_varycon}). This represents a
strong hint that 
current-induced corrections are for the $\nu=2$-QHD at least as
important as for the $\nu=1$-QHD. This aspect is not considered in
calculations relying on an interpolation of the type of Eq.\
(\ref{interp}) which neglects the influence of higher Landau levels.

\section{Vorticity-dependent XC energy densities}
\label{v2e}

In this Section we will deal with the vorticity dependence of XC
energy densities extracted from the XC scalar and
vector potentials above. The two parts comprise the methodology and the
results for the four-electron $\nu=2$ system.

\subsection{Calculation of XC energy densities}

The steps for extracting XC energy densities from XC potentials are
analogous to 
Ref.\ [\onlinecite{wensauer03d}] but generalized with respect to the
vorticity dependence. Eqs.\ (\ref{vxcdef}) and (\ref{axcdef})
provide the general relation between XC functional
and XC potentials.
To construct a relation between XC potentials and XC energy
densities we apply the LVSDA (\ref{lvsdaeq})
and obtain
\begin{eqnarray}
&&eA_{{\rm XC},\varphi}(r)=-\frac{1}{n(r)}\frac{\partial}{\partial r}
\left(n(r) \frac{\partial \varepsilon_{\rm XC}}{\partial v}(r)
\right)
\\
&&V_{{\rm XC},\sigma}(r)=\varepsilon_{\rm XC}(r)
+n(r)\frac{\partial \varepsilon_{\rm XC}}{\partial n_\sigma}(r)
-eA_{{\rm XC},\varphi}(r)\frac{j_{{p},\varphi}(r)}{n(r)}.
\end{eqnarray}
We solve for $\frac{\partial \varepsilon_{\rm XC}}{\partial
  v}(r)$ and $\frac{\partial \varepsilon_{\rm XC}}{\partial
  n_\sigma}(r)$
\begin{eqnarray}
\frac{\partial\varepsilon_{\rm XC}}{\partial v}(r)
&=&\frac{1}{n(r)}
\left[\int\limits_r^\infty {\rm d}r^{\prime\prime}n(r^{\prime\prime})
eA_{{\rm XC},\varphi}(r^{\prime\prime})\right]
\label{axc1}\\
\frac{\partial \varepsilon_{\rm XC}}{\partial n_\sigma}(r)
&=&\frac{1}{n(r)}\left(-\varepsilon_{\rm XC}(r)+V_{{\rm XC},\sigma}(r)
+eA_{{\rm XC},\varphi}(r)\frac{j_{{p},\varphi}(r)}{n(r)}\right)
\label{vxc1}
\end{eqnarray}
considering the (physical) boundary condition
\begin{equation}
\lim_{r\rightarrow\infty}n(r)\frac{\partial\varepsilon_{\rm
XC}}{\partial v}(r)=0.
\end{equation}
The results are substituted in the derivative of the XC
energy density as a function of the radius
\begin{equation}
\frac{\partial \varepsilon_{\rm XC}(r)}{\partial r}=
\sum_\sigma \frac{\partial \varepsilon_{\rm XC}}{\partial n_\sigma}
\frac{\partial n_\sigma}{\partial r}+\frac{\partial \varepsilon_{\rm
    XC}}{\partial v} \frac{\partial v}{\partial r}.
\end{equation}
After some transformations we arrive at a linear differential
equation (DEQ)
\begin{equation}
\frac{\partial \varepsilon_{\rm XC}(r)}{\partial r}+
\frac{\partial\log n}{\partial r}(r)\varepsilon_{\rm XC}(r)=I(r)
\label{deq}
\end{equation}
with the inhomogeneity
\begin{eqnarray}
I(r)&=&\frac{1}{n(r)}\sum_\sigma \left(V_{{\rm
XC},\sigma}(r)+c_\sigma\right) 
\frac{\partial n_\sigma}{\partial r}(r)
+eA_{{\rm XC},\varphi}(r)\frac{j_{{p},\varphi}(r)}{n(r)}
\frac{\partial\log n}{\partial r}(r)
\nonumber \\
&&+\frac{1}{n(r)}\frac{\partial v}{\partial r}(r)
\left[\int\limits_r^\infty {\rm d}r^{\prime\prime}n(r^{\prime\prime})
eA_{{\rm XC},\varphi}(r^{\prime\prime})\right]
\label{inh}
\end{eqnarray}
which contains all the information about the dependence on the spin
densities and the vorticity. 
In Eq.\ (\ref{inh}) we take into account that the gauge constants
$c_\sigma$ of the scalar potentials are not known from the previous
step. They will be calculated later.
The solution of the homogeneous part of DEQ (\ref{deq})
is given by
$\varepsilon_{\rm XC}^{\rm hom}(r)=\alpha/n(r)$,
a special solution can be calculated using the ansatz
$\varepsilon_{\rm XC}^{\rm spec}(r)=\beta(r)/n(r)$.
The function $\beta(r)$ follows from an elementary DEQ
\begin{equation}
\frac{\partial \beta}{\partial r}(r)=n(r)I(r){\text ,}
\end{equation}
whose solution is
\begin{equation}
\beta (r)=\int\limits_0^r{\rm
d}r^\prime\,n(r^\prime)I(r^\prime)-\beta(0).
\end{equation}
Thus the general solution of DEQ (\ref{deq}) is
\begin{equation}
\varepsilon_{\rm XC}(r)=\varepsilon_{\rm XC}^{\rm
hom}(r)+\varepsilon_{\rm XC}^{\rm
spec}(r)=\frac{\alpha-\beta(0)}{n(r)} 
+\frac{1}{n(r)}\int\limits_0^r{\rm
d}r^\prime\,n(r^\prime)I(r^\prime).
\label{gen}
\end{equation}
After substituting the inhomogeneity the result reads
\begin{eqnarray}
\varepsilon_{\rm XC}(r)&=&\frac{\alpha-\beta(0)
-n_\sigma(0)\sum_\sigma c_\sigma}{n(r)}+\sum_\sigma c_\sigma\frac{n_\sigma(r)}{n(r)}
+\frac{1}{n(r)}
\int\limits_0^r{\rm d}r^\prime
{\Bigg\lbrace}\sum_\sigma V_{{\rm XC},\sigma}(r^\prime) 
\frac{\partial n_\sigma}{\partial r^\prime}(r^\prime)
\nonumber\\
&&+eA_{{\rm XC},\varphi}(r^\prime){j_{{p},\varphi}(r^\prime)}
\frac{\partial\log n}{\partial r^\prime}(r^\prime)
+\frac{\partial v}{\partial r^\prime}(r^\prime)
\left[\int\limits_{r^\prime}^\infty {\rm d}r^{\prime\prime}n(r^{\prime\prime})
eA_{{\rm XC},\varphi}(r^{\prime\prime})\right]
{\Bigg\rbrace}.
\end{eqnarray}
With the modulus of the XC energy being finite in LVSDA
\begin{equation}
|E_{\rm XC}[n_\sigma,v]|\approx \left|\int {\rm d}{\bf r}
 \,n(r)\varepsilon_{\rm XC}(r)\right|= \int {\rm d}{\bf r}
 \,n(r)|\varepsilon_{\rm XC}(r)| <\infty
\end{equation}
we choose $\alpha-\beta(0)-n_\sigma(0)\sum_\sigma c_\sigma=0$ thus
avoiding any divergent contributions.
Consequently the analytical solution of (\ref{deq}) satisfying the
physical boundary conditions is
\begin{eqnarray}
\varepsilon_{\rm XC}(r)&=&\sum_\sigma c_\sigma\frac{n_\sigma(r)}{n(r)}
+ \frac{1}{n(r)}
\int\limits_0^r{\rm d}r^\prime
{\Bigg\lbrace}\sum_\sigma  V_{{\rm XC},\sigma}(r^\prime) 
\frac{\partial n_\sigma}{\partial r^\prime}(r^\prime)
+eA_{{\rm XC},\varphi}(r^\prime){j_{{p},\varphi}(r^\prime)}
\frac{\partial\log n}{\partial r^\prime}(r^\prime)
\nonumber\\
&&+\frac{\partial v}{\partial r^\prime}(r^\prime)
\left[\int\limits_{r^\prime}^\infty {\rm d}r^{\prime\prime}n(r^{\prime\prime})
eA_{{\rm XC},\varphi}(r^{\prime\prime})\right]
{\Bigg\rbrace}.
\label{exc2}
\end{eqnarray}
The last step is the calculation of the gauge constants $c_\sigma$
of the scalar potentials. One condition which has not been used
yet is the agreement of the DFT GS energy and the exact GS energy.
If we eliminate $V_{{\rm XC,}\sigma}({\bf r})$ in Eq.\ (\ref{Eformel})
and write it in a modified form
\begin{equation}
E_{GZ}=\sum_{j,\sigma}\gamma_{j,\sigma}
\int {\rm d}{\bf r}\,\varphi_{j,\sigma}^\ast({\bf r})
\left(\frac{1}{2m^\ast}\left({\bf p}+e{\bf A}({\bf
      r})\right)^2+V_\sigma({\bf r})\right) 
\varphi_{j,\sigma}({\bf r})+\frac{1}{2}\int {\rm d}{\bf r}\,V_{\rm H}({\bf r})n({\bf r})
+E_{\rm XC}[n_\sigma,v]
\label{ksgs}
\end{equation}
we can calculate the (exact!) XC energy $E_{\rm XC}[n_\sigma,v]$ for
GS densities: the KS wavefunctions are known from the selfconsistent
solution of the KS equations. Thus the expectation
values $\int {\rm d}{\bf r}\,\varphi_{j,\sigma}^\ast({\bf r})
\left(\frac{1}{2m^\ast}\left({\bf p}+e{\bf A}({\bf
      r})\right)^2+V_\sigma({\bf r})\right) 
\varphi_{j,\sigma}({\bf r})$ of the non-interacting system can be
calculated. The Coulomb energy and the GS-energy are uniquely
determined by the interacting system.
On the other hand the XC energy in LVSDA is given by
\begin{eqnarray}
E_{\rm XC}[n_\sigma,{\bf v}]&\approx&
\int{\rm d}{\bf r}\,n(r)
\varepsilon_{\rm XC}(r)\nonumber\\
&=&\sum_\sigma c_\sigma N_\sigma+\int\limits_{r=0}^{\infty}{\rm d}r
\int\limits_0^r{\rm d}r^\prime
{\Bigg\lbrace}\sum_\sigma V_{{\rm XC},\sigma}(r^\prime)
\frac{\partial n_\sigma}{\partial r^\prime}(r^\prime)
+eA_{{\rm XC},\varphi}(r^\prime)
\times\nonumber\\
&&\times
{j_{{p},\varphi}(r^\prime)}
\frac{\partial\log n}{\partial r^\prime}(r^\prime)
+\frac{\partial v}{\partial r^\prime}(r^\prime)
\left[\int\limits_{r^\prime}^\infty {\rm d}r^{\prime\prime}n(r^{\prime\prime})
eA_{{\rm XC},\varphi}(r^{\prime\prime})\right]
{\Bigg\rbrace}
\label{ksgs2}
\end{eqnarray}
what makes $\sum_\sigma c_\sigma N_\sigma$ accessible.
In the case of unpolarized systems $c:=c_\uparrow=c_\downarrow$, 
and for full polarization the constant $c_\sigma$ of the unoccupied spin
direction is irrelevant (as the corresponding
$N_\sigma=0$). Consequently, the results for the XC 
energy density are unique. For partially  
polarized systems uniqueness of the results can be achieved by
additionally demanding asymptotic agreement of
$V_{\rm XC,\sigma}(r)$ for finite systems\cite{capelle01}  
\begin{equation}
\lim\limits_{r\to\infty}V_{\rm
  XC,\uparrow}(r)=\lim\limits_{r\to\infty}V_{\rm XC,\downarrow}(r). 
\end{equation}

As a result of the inversion of the LVSDA formalism we obtain the
XC energy density as a function of the radius. After eliminating the
radius using the spin densities $n_\sigma(r)$ and vorticity $v(r)$
or alternatively $r_s(r)$, $\xi(r)$, and $v(r)$ we arrive at the
representation of the XC energy density as a function of $(r_s,\xi,v)$. 

\subsection{Results for vorticity-dependent XC energy densities}

In this Section we summarize the numerical results for the extracted
XC energy densities. Because of the high contribution of
self-interaction in the two-electron system only the four-electron
$\nu=2$-QHD can be used for the derivation of vorticity-dependent XC
energy densities. After solving the DEQ (\ref{deq}) for different
confinement energies $\omega_0$ we get a set of $\varepsilon_{\rm XC}$
for different $r_s$, $v$, and $\xi=0$ (no polarization). The results
for different vorticities are shown in Fig.~\ref{xc_energy} as a
function of $r_s$ together with the pure TC\cite{tanatar89}
parameterization, TC combined with
LWM\cite{tanatar89,levesque84,rasolt92}, and TC combined with
FO\cite{tanatar89,fano88,rasolt92}. Whereas the LWM and FO
parameterizations do not deviate very much from the pure TC data
which does not consider any current-induced effects, the extracted XC
energy densities exhibit a pronounced dependence on the
vorticity. In particular, the curve of extracted $\varepsilon_{\rm
  XC}$ starts for small $r_s$ below the reference curves, crosses it
and finally lies above it. Moreover, the crossing wanders to larger $r_s$
if we decrease the vorticity (see Fig.~\ref{xc_energy}(a)-(f)). 

Although the interpretation of these large deviations from the
standard parameterizations is difficult as the data set is very
small our results give rise to the question if the vorticity
dependence of XC energy densities can be treated in the frame of
LVSDA. There might be also non-local effects from paramagnetic
currents which cannot be considered in this approach. However, in any
case it is necessary to reexamine the vorticity-dependence of
$\varepsilon_{\rm XC}$ with focus on Landau level mixing and
arbitrary densities.

\section{Summary}

In this paper we presented a new approach to study scalar XC
potentials and XC vector potentials of current-carrying
ground states in axially symmetric, parabolic quantum dots. Starting
from exact ground states densities and paramagnetic current
densities we analytically calculated
the XC potentials of the 
two-electron $\nu=1$-droplet and the four-electron $\nu=2$-droplet
and studied them as a function of the strength of the confinement
potential. In both systems the XC vector potentials are of the same
order of magnitude which shows the importance of this
current-induced effect even for higher filling factors. This result
is not taken into account in the interpolation introduced by Rasolt and
Perrot\cite{rasolt92}. 

The second step involved the extraction of vorticity-dependent XC
energy densities. The effect of the vorticity is far more
pronounced than predicted by Rasolt and Perrot\cite{rasolt92},
Levesque, Weis, and 
MacDonald\cite{levesque84} and Fano and Ortolani\cite{fano88}. A
possible reason for 
this behavior is the assumption that the influence of current
densities is only relevant for strong magnetic fields when the local
filling factor $f$ is less than one. This is the restriction which
enters the derivations of the parameterizations in
literature. However, our results point out that a new approach to
current- or vorticity-dependent XC energy densities is required which also
comprise the effects of Landau level mixing.

\section{Acknowledgement}
A.\ W.\ thanks the the RRZE Erlangen and the DFG (Ro 522/19-1).

\newpage

\noindent
\unitlength1cm
\begin{figure}
\caption{Analytical results for the two-electron $\nu=1$-QHD
  with confinement energy $2/3\,{\rm Ry}$. Figs.\ (a)-(h) depict the
  GS density, the paramagnetic GS density, the vorticity, the radial
  KS wavefunctions, the effective scalar potential, the Hartree
  potential, the scalar XC potential, and the XC vector potential.
\label{2el_ana}
}
\end{figure}

\noindent
\unitlength1cm

\noindent
\unitlength1cm

\noindent
\unitlength1cm

\noindent
\unitlength1cm
\begin{figure}
\caption{Vorticity-dependent XC energy densities extracted from the
  four-electron $\nu=2$-QHD. (a)-(f) show the XC energy densities as
  a function of $r_s$ for different vorticities. 
\label{xc_energy}
}
\end{figure}

\end{document}